\documentclass{article}

\usepackage{graphicx}
\usepackage{amsmath}
\usepackage{amssymb}

\begin{document}

\noindent {\bf Identification of the Thermal Conductance of a hidden barrier from outer thermal data } \\ 
\noindent {G. Inglese, R. Olmi} \\ \\

\noindent {\it Abstract} Hidden defects affecting the interface in a composite slab are evaluated from thermal data collected on the upper side of the specimen.  First we restrict the problem to the upper component of the  object. Then we investigate heat transfer through the inaccessible interface by means of Thin Plate Approximation. Finally, Fast Fourier Transform is used to filter data. In this way  we obtain a reliable reconstruction of simulated flaws in thermal contact conductance corresponding to appreciable defects of the interface.

\section{Introduction}

\begin{figure}[!ht]
\begin{center}
\includegraphics[width=0.8\textwidth]{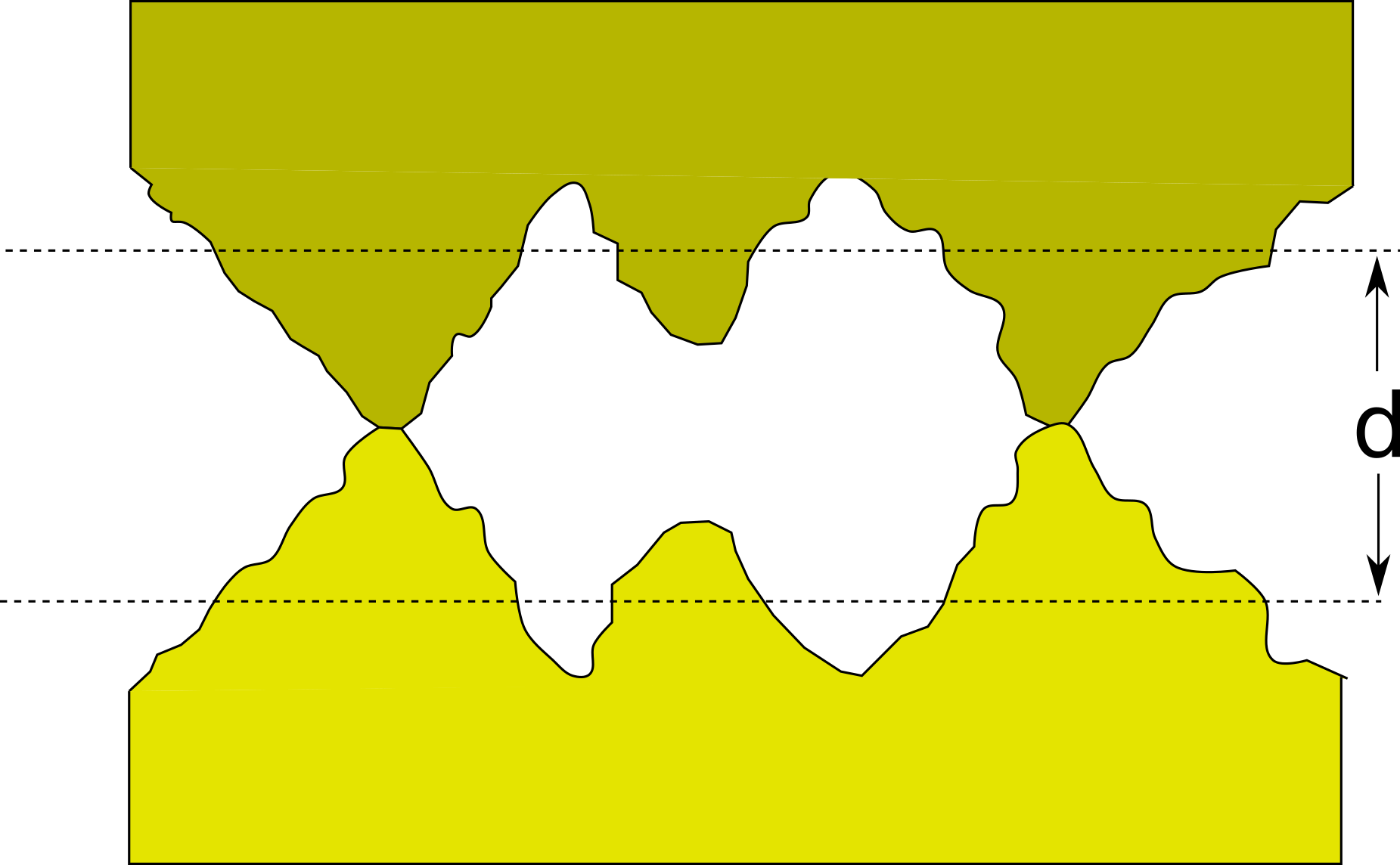}
\caption{Contact between two real solid surfaces}
\label{disegno}
\end{center}
\end{figure}

Consider a composite body made up of two slabs of different materials in close thermal contact.
Since the contact surfaces are rough on a microscopic level, thermal contact is 
always {\it imperfect} (see for example \cite{IDBL03}-Section 3.1 and \cite{M74}). With reference to figure \ref{disegno}, 
taking into account that the true contact area is a small portion of the apparent contact area,  the  slabs are separated by an interface of average width $d$, filled up with air. 
The present work deals with  the nondestructive evaluation of deviations of the interface width from a given average value.  Since $\frac{\kappa_{air}}{d}$ defines the {\it thermal conductance} between the slabs, a thermal model of the composite material can be implemented to evaluate local variations of the width by applying a controlled heat flux  and collecting a sequence of temperature maps on the top side of the body (Active Thermography \cite{Ma01}).\\
  
 A recent effective approach to the solution of this problem, based on Reciprocity Functional technique, is described in \cite{ACOA16}.  Here, a perturbative tool like Thin Plate Approximation (TPA)(see for example \cite{IOS17}) is used in alternative (see section \ref{sec:TPA}). The idea of using TPA comes from the general assumption (see \cite{Ma01} Sect 9.2.1) that thermography is effective in  detection of subsurface anomalies. Here, TPA is expected to work since the Biot number of the upper slab is $\le .1$ so that it can be regarded as a {\it thermally thin} domain (see \cite{IDBL03} Section 5.2).\\

 
 
\section{Geometry of the specimen}\label{sec:geome}

Consider the composite domain $\Omega_A=\Omega_+ \cup A \cup \Omega_-$ where
$$\Omega_+=\{(x,y,z)\phantom{a} s.t.\phantom{a} x,y \in (-L,L) \phantom{a} a^+ > z > \epsilon g_+(x,y) \}$$
$$\Omega_-=\{(x,y,z)\phantom{a}s.t.\phantom{a} x,y \in (-L,L) \phantom{a} -a ^-< z < - \epsilon g_-(x,y) \}$$
with $g^+$ and $g^-$ continuous, possibly non differentiable, functions ranging in $(0,1)$ with $\epsilon << \min \{a^-,a^+\}$.


We stress that $\Omega_+$ and $\Omega_-$ are made of different materials each characterized by density $\rho_\pm$, specific heat $c_\pm$ and thermal conductivity $\kappa_\pm$. The third, irregular, thin slab is 
\begin{equation}
A=\{(x,y,z)\phantom{a}s.t.\phantom{a} x,y \in (-L,L) \phantom{a}- \epsilon g_-(x,y) < z < \epsilon g_+(x,y)  \}
\end{equation}
and $\rho_a$, $c_a$ and $\kappa_a$ are its physical parameters.
Let  $\gamma=\frac{a}{L}$ a dimensionless parameter which represents the geometrical "thinness" of the slab $\Omega_+$.\\

The domain $A$ has variable thickness $\epsilon(g_+(x,y)-g_-(x,y))$ and it is assumed to be  filled up with air, whose conductivity $\kappa_a=0.002587$ $\mathrm{W m^{-1} K^{-1}}$ is much lower than  $\kappa_{\pm}$.  

It means that $A$ opposes to heat transfer from $\Omega^+$ to $\Omega^{-}$ and the corresponding  Thermal Conductance is
\begin{equation}
H(x,y)= \frac{\kappa_a}{\epsilon(g_+(x,y)-g_-(x,y))}.
\end{equation}

\section{Modeling the solid interface $A$ by means of Robin boundary conditions on the two sides of a plane. Imperfect contact.}

The temperature of  $\Omega_A$ depends on the physical characteristics of $A$.
We account for the different conductivities $\kappa_-$, $\kappa_a$, $\kappa_+$  in $\Omega_A$ by imposing continuity of  temperature  and heat flux for $z=\epsilon g_+(x,y)$ and $z=-\epsilon g_-(x,y)$ as transmission conditions for the heat conduction equation in $\Omega_A$. 
The functions  $g_+$ and $g_-$ are very irregular at a microscopic scale but, if their values are normally distributed in a small neighborhood of the mean values $\overline{g}_- $ and $\overline{g}_+ $,
the set $A$ can be successfully approximated by the parallelepiped $\bar{A}=\{(x,y,z)\phantom{a}s.t.\phantom{a} x,y \in (-L,L)  ,\phantom{a}- \epsilon \overline{g}_-  < z < \epsilon \overline{g}_+  \}$. 

The task of solving the heat equation in $\Omega_A$  becomes much simpler if we consider $\bar{A}$ instead of $A$.  The interface behaves like a controlled heat exchanger  between $\Omega_+$ and $\Omega_-$ of conductance $ H= \frac{\kappa_a}{\epsilon(\overline{g}_+-\overline{g}_-)}$. A further simplification consists in squeezing $\bar{A}$ to the plane $z=0$ and assigning the Robin boundary conditions (imperfect contact \cite{Me43})
\begin{equation}\label{Robconsth}
k_\pm u^\pm_n (x,y,0^\pm)+ H(u^\pm(x,y,0^\pm)-u^\mp(x,y,0^\mp))=0.
\end{equation}
while the heat equation is considered in the two domains $\Omega_+=(-L,L)\times(-L,L)\times(0,a^+)$ and $\Omega_-=(-L,L)\times(-L,L)\times(-a^-,0)$.\\

The presence of an anomaly in the interface, corresponds to a change in the heat transfer from $\Omega_+$ to $\Omega_-$. Hence, in analogy with (\ref{Robconsth}), we have
\begin{equation}\label{Robin}
k_\pm u^\pm_n (x,y,0^\pm)+H(x,y)\left(u^\pm(x,y,0^\pm)-u^\mp(x,y,0^\mp)\right)=0.
\end{equation}
where
\begin{equation}
H(x,y)= \frac{\kappa_a}{\epsilon(g_+(x,y)-g_-(x,y))}.
\end{equation}
These boundary conditions describe a non constant imperfect contact (see \cite{BK87} \cite{InOl17}).  The pointwise evaluation of the unknown function $H(x,y)$  is the {\it main goal} of the present work.\\

\noindent {\bf Remark} We are assuming that variations of $H$ in time take place in a much longer  interval than $(0,t_{max})$. For this reason,  $H$ does not depend on $t$.\\

\section{The direct model}\label{sec:direct}

Let $u^+$ be the solution of 
\begin{equation}\label{equpiu}
\rho_+ c_+ u_t= \kappa_+ \Delta u
\end{equation}
in $(-L,L)\times (-L,L)\times(0,a)\times (0,t_{max}]$ with initial condition 
\begin{equation}\label{init}
u^+(x,y,z,0)=0 .
\end{equation}
Boundary data are  
\begin{equation}\label{bound01}
u^+_x(-L,y,z,t)=u^+_x(L,y,z,t)=u^+_y(x,-L,z,t)=u^+_y(x,L,z,t)=0
\end{equation}
and 
\begin{equation}\label{bound02}
\kappa_+ u^+_z(x,y,a,t)= \Phi(x,y)\chi_{(0,t_S)}(t)
\end{equation}
i.e. a source of power $\Phi$ is ON for $t_S$ seconds ($t_S<t_{max}$ while the index $S$ is for "source").\

Let $u^-$ be the solution of 
\begin{equation}
\rho_- c_- u^-_t= \kappa_- \Delta u^-
\end{equation}
in $(-L,L)\times (-L,L)\times(-a,0)\times (0,t_{max}]$ with initial condition $u^-(x,y,z,0)=0$. Boundary data  are $u^-_x(-L,z,t)=u^-_x(L,z,t)=u^-_y(x,-L,z,t)=u^-_y(x,L,z,t)=0$ and $u^-(x,y,-a,t)=0$.\\

Transmission conditions 
$$\lim_{\epsilon \to 0} u^+(x,y,\epsilon,t)=\lim_{\epsilon \to 0} u^-(x,y,-\epsilon,t)$$
 and 
 $$\kappa_+ \lim_{\epsilon \to 0} \frac{\partial u^+}{\partial z}(x,y,\epsilon,t)=\kappa_- \lim_{\epsilon \to 0} \frac{\partial u^-}{\partial z}(x,y,-\epsilon,t)$$
  holds and can be written in terms of boundary data  on the two sides of the interface $z=0$. More precisely
\begin{equation}\label{Robupm}
-\kappa_+ u^+_z+ H(x,y)(u^+(x,y,0^+,t)-u^-(x,y,0^-,t))=0
\end{equation} 
\begin{equation}
\kappa_- u^-_z+H(x,y)(u^-(x,y,0^-,t)-u^+(x,y,0^+,t))=0.
\end{equation}
We stress that  solutions $u^+$ and $u^-$ depend  on the dimensionless parameter $\gamma=\frac{a}{L}$ introduced in section \ref{sec:geome}. 
 \\ \\
\subsection{From transmission conditions to Robin Boundary Conditions}\label{sec:TrasRob}

\begin{figure}[!ht]
\begin{center}
\includegraphics[width=0.8\textwidth]{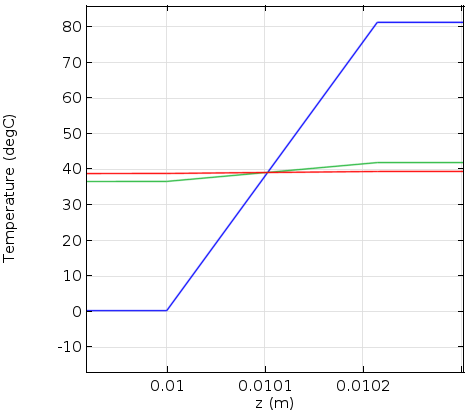}
\caption{Dependence of the temperature gap on the  heat exchange coefficient at the interface (see text)}
\label{transmission}
\end{center}
\end{figure}

It is well known that, when $\kappa_+=\kappa_-$, the midpoint of the thermal jump due to imperfect contact is the value of the background temperature . It means that transmission conditions can be changed in ordinary boundary conditions on the two sides of the interface (see for example \cite{IOS19}). Here, it is  
$$u^+(x,y,0^+,t)+u^-(x,y,0^-,t) = 2 u^0(x,y,0,t) -R(a,H,\kappa_+,\kappa_-,\alpha_+,\alpha_-;x,y,t)$$
for all $H$, $x,y$ and $t$. When $H=0$ (insulating interface) and $H=\infty$ (perfect contact) we have $R=0$. $R>0$ elsewhere.
There is numerical evidence that $R$ is negligible when parameters are in the range we are dealing with. Figure 
\ref{transmission} shows the temperature gap across the interface for a heat exchange coefficient changing from 0 (blue curve) to $\infty$ (red curve), compared to that relative to $H = 1000$ (green curve). Moreover, in an analogous simpler problem for composite regions (\cite{CJ59} sect 14.6)  in which the temperature is known explicitly, straightforward calculations lead to the following formula for "large" $H$:
\begin{equation}\label{esti}
R(a,H,\kappa_+,\kappa_-,\alpha_+,\alpha_-;x,y,t) \approx G_0 \frac{\kappa_+\sqrt{\alpha_-}-\kappa_-\sqrt{\alpha_+}}{\kappa_+\sqrt{\alpha_-}+\kappa_-\sqrt{\alpha_+}}\frac{a}{2t\sqrt{\alpha_+} S H}
\end{equation}
where $G_0(x,y,0,t; 0,0,a,0)$ is the Green function of a Continuous Plane Source and the constant $S$ is
\begin{equation}
    S=\frac{\kappa_+\sqrt{\alpha_-}+\kappa_-\sqrt{\alpha_+}}{\kappa_+\kappa_-}.
\end{equation}

Assumed the smallness of $R$ (confirmed by numerical computations and by the analytical estimate (\ref{esti}), interface boundary conditions can be written approximately as 
\begin{equation}\label{Robu0}
-\kappa_+ u^+_z+2 H(x,y)(u^+(x,y,0^+,t)-u^0(x,y,0,t))=0
\end{equation} 
\begin{equation}
\kappa_- u^-_z+2 H(x,y)(u^-(x,y,0^-,t)-u^0(x,y,0,t))=0.
\end{equation}\\
The IBVP (\ref{equpiu})(\ref{init})(\ref{bound01})(\ref{bound02})(\ref{Robu0}) is now the direct model underlying the inverse problem of evaluating  $H$.  Observe that the problem has been restricted to the upper slab  $\Omega_+$.


\section{Dimensionless variables and Thin Plate Approximation}\label{sec:TPA}

The dimensionless parameter $\gamma=\frac{a}{L}$ gives a measure of how much $\Omega_+$ is "geometrically thin". As observed in the introduction, $\Omega_+$ is considered "thermally thin" when $\frac{aH}{\kappa} < .1$. We introduce the set of dimensionless variables $\zeta=1-\frac{z}{\gamma L}$, $\xi=\frac{x}{L}$, $\eta=\frac{y}{L}$ and $\tau=t \frac{\alpha_+}{ L^2}$ (we recall that the numbers $\alpha_\pm=\frac{\kappa_\pm}{\rho_\pm c_\pm}$ are the diffusivities of upper and lower slabs respectively) and define
\begin{equation}\label{vvsu}
v^{+,\gamma}(\xi,\eta,\zeta,\tau) \equiv u^+(L\xi,L\eta,L\gamma(1-\zeta),\frac{L^2}{\alpha_+}\tau;\gamma)
\end{equation} 
so that,
the heat equation for $\zeta>0$ (former upper slab) becomes
\begin{equation}\label{equat}
 \gamma^2 v_\tau = \gamma^2 (v_{\xi \xi}+v_{\eta \eta})+ v_{\zeta \zeta}
\end{equation}
for $(\xi,\eta,\zeta) \in (0,1) \times 0,1) \times (0,1)$ and $\tau \in (0,t_{max} \frac{\alpha_+}{L^2})$. Observe that in (\ref{vvsu}) we use a notation that points out the explicit dependence of $v^{+,\gamma}$ on $\gamma$.  Variables $\xi$, $\eta$, $\zeta$ and $\tau$ are slightly different from usual ones (see for example \cite{LS88} Sect 4.1) because we need to normalize  $\zeta$ by means of the parameter $\gamma$ which, consequently, is present in (\ref{equat}). \\

Thanks to the substitution of (\ref{Robupm}) with (\ref{Robu0}) , we can limit ourselves to the upper slab $\zeta \in (0,1)$ . Furthermore, we scale suitably the heat flux
$$\gamma\phi=\Phi$$
and thermal conductances
$$\gamma h(\xi,\eta)=H(L\xi,L\eta)$$
$$\gamma h_+=H_+$$
obtaining the boundary conditions
\begin{equation}\label{etazero}
-\frac{\kappa_+}{L} v_\zeta(\xi,\eta,0,\tau)+\gamma^2 h_+ v(\xi,\eta,0,\tau) = \gamma^2 \phi(L\xi,L\eta) \chi_{(0, \frac{\alpha t_S}{L^2})}(\tau)
\end{equation}

\begin{equation}\label{etauno}
\frac{\kappa_+}{L} v_\zeta(\xi,\eta,1,\tau)+\gamma^2 2h(L\xi,L\eta) (v(\xi,\eta,1,\tau)-u^0(L\xi,L\eta,0,\frac{L^2}{\alpha_+}\tau)) = 0
\end{equation}
and
\begin{equation}
v_\xi(-1,\eta,\zeta,\tau)= v_\xi(1,\eta,\zeta,\tau)=0
\end{equation}
\begin{equation}
v_\eta(\xi,-1,\zeta,\tau)= v_\eta(\xi,1,\zeta,\tau)=0.
\end{equation}

\subsection{Formal expansion of $v^{+,\gamma}$}
Consider the formal expansions
\begin{equation}
v^{+,\gamma}(\xi,\eta,\zeta,\tau)=v_0(\xi,\eta,\zeta,\tau)+\gamma v_1(\xi,\eta,\zeta,\tau)+\gamma^2 v_2(\xi,\eta,\zeta,\tau)+.... \equiv u^+(L\xi,L\gamma(1-\zeta),\frac{L^2}{\alpha_+}\tau;\gamma)
\end{equation}
and
\begin{equation}\label{hexpa}
h(\xi,\eta)=h_0(\xi,\eta,\tau)+\gamma h_1(\xi,\eta,\tau)+... \equiv \frac{H_{eff}(x,y)}{\gamma}
\end{equation}
It is remarkable that, though $h$ is independent on $\tau$ (at least in the time scale at hand), its partial sums in the expansion above are dependent on $\tau$ by construction.\\

We observe that  
\begin{equation}\label{v0}
v_0(\xi,\eta,\zeta,\tau)=\lim_{\gamma \to 0} u^+(L\xi,L\eta,L\gamma(1-\zeta),\frac{L^2}{\alpha_+}\tau;\gamma).
\end{equation}
We claim that the limit in (\ref{v0}) exists as we can see by generalizing the  one dimensional case with $\phi(x,y)=\phi_0$ and constant $h$. In this special case, it is  $u^+(z,t)=\frac{\gamma \phi_0}{\kappa}z+\frac{\phi_0}{2 h }+O(\gamma^2)e^{-\beta t}$ so that we extrapolate that 
$$\lim_{\gamma \to 0} u^+(L\xi,L\eta,L\gamma(1-\zeta),\frac{L^2}{\alpha_+}\tau;\gamma)=\frac{\phi(x,y)\chi_{(0,t_S)}(t)}{2h(x,y)}$$
 for $t \in (0,t_S)$. \\

As for the other coefficients, we have that $v_N(\xi,\eta,\zeta,\tau)=\lim_{\gamma \to 0} \frac{1}{N!} \frac{\partial^N  v^{+,\gamma}}{\partial \gamma^N}$. In particular:
\begin{equation}\label{v1}
v_1(\xi,\eta,\zeta,\tau)= \lim_{\gamma \to 0}(-u^+_z(L\xi,L\eta,L\gamma(1-\zeta),\frac{L^2}{\alpha_+}\tau;\gamma)L(1-\zeta)+\frac{\partial u^+}{\partial \gamma})
\end{equation}

It is remarkable that in the  one dimensional case it is $\frac{\partial u^+}{\partial \gamma}=\frac{\Phi_0}{\kappa}((1-\zeta)\gamma L+O(\gamma)e^{-\beta t}$ (see \cite{CJ59}). Hence, it is easy to see that $v_1(\xi,\eta,\zeta,\tau) \to 0 $ for $\gamma \to 0$. \\

Plugging the expansion of $v^{+,\gamma}$ in (\ref{equat}), (\ref{etazero}) and (\ref{etauno}) we have
\begin{equation}
v_{0\zeta\zeta}=v_{1\zeta\zeta}=0
\end{equation} 
with $v_{0\zeta}=v_{1\zeta}=0$ i.e. $\frac{\partial (v_0+\gamma v_1)}{\partial \zeta}= 0$.
As suggested by (\ref{v0}) and (\ref{v1}), we have
\begin{equation}
v_0(\xi,\eta,\zeta,\tau) \equiv v_0(\xi,\eta,\tau) \approx u^+(x,y,a,t) 
\end{equation} 
(the symbol $\approx$ means that $u^+(x,y,a,t)$ is taken in practice on the top side of a thin plate of thiockness $a>0$) and
\begin{equation}
v_1(\xi,\eta,\zeta,\tau)=\lim_{\gamma \to 0} \frac{\phi_0}{\kappa}(1-\zeta)\gamma L=0.
\end{equation}\\

The order zero approximation of $h$ can be obtained from second order relations. More precisely,
\begin{equation}\label{order2}
v_{0\tau}=v_{0\xi\xi}+v_{0\eta\eta}+v_{2\zeta\zeta}
\end{equation}
\begin{equation}\label{etazero0}
-\frac{\kappa_+}{L} v_{2\zeta}(\xi,\eta,0,\tau)+h_+ v_0(\xi,\eta,\tau) = \phi(L\xi,L\eta) \chi_{(0, \frac{\alpha t_S}{L^2})}(\tau)
\end{equation}
\begin{equation}\label{etauno0}
\frac{\kappa_+}{L} v_{2\zeta}(\xi,\eta,1,\tau)+2h_0 (\xi,\eta,\tau)(v_0(\xi,\eta,\tau)-u^0(L\xi,L\eta,0, \frac{L^2}{\alpha_+}\tau)) = 0
\end{equation}
so that
\begin{equation}
2h_0(\xi,\eta,\tau) = \frac{-h_+v_0+\phi+\frac{\kappa_+}{L}(v_{0\xi\xi}+v_{0\eta\eta}-v_{0\tau})}{v_0-u^0(x,y,0, t)}.
\end{equation}
Choosing a time value $\bar{t}$ which corresponds to a good approximation of the "ideal" value of $H$ (i.e. its value in absence of conduction flaws)  and coming back to  variables $(x,y,z,t)$, we have
\begin{equation}\label{h_formula}
2H(x,y) \approx \frac{\Phi(x,y)\chi_{(0,t_S)}(\bar{t})+a(\kappa_+T_{xx}+\kappa_+T_{yy}-\rho_+ c_+ T_{t})-a\frac{h_+}{L}T}{T-u^0(x,y,0, t)}
\end{equation}
where $T=u(x,y,a,\bar{t})$ is the temperature of the accessible side $z=a$ at time $\bar{t}$. Recall that, in experimental real life situations, our knowledge of $T$ come from a sequence of measurements taken by means of an infrared camera. In numerical simulations $T$ is computed solving the direct model with a finite elements code. \\

\noindent{\it Remark.} When the unknown parameter $H$ is a function of two variables, the computation of higher order term in expansion (\ref{hexpa}) is very hard. A strong smoothing procedure would be required in order to perform reliable computation of fourth or sixth order partial derivative of data $T$. At present we have not yet solved this problem.


\section{Numerical computations}\label{sec:numer}


Simulation of experimental data collection requires the numerical solution of the direct problem described in section \ref{sec:direct}. 
Numerical values of parameters (in MKS units) are:\\

\noindent
Upper slab
\begin{itemize}
\item $\kappa_{+} = 54$
\item $\rho_{+} = 7870$
\item $c_{+} = 486$
\end{itemize}
Lower slab
\begin{itemize}
\item $\kappa_{-}$ = 14
\item $\rho_{-}$ = 8000
\item $c_{-}$ = 500
\end{itemize}

\noindent
Geometrical parameters:
\begin{itemize}
\item $d_0 = 10^{-5}\ m$ (average contact thickness)
\item $a = 10^{-2}\ m$ (thickness of a single slab)
\item $L = 0.1\ m$ (slab side)
\end{itemize}

\noindent The upper surface is uniformly illuminated with a source (spotlight)  having a power density 
\begin{itemize}
\item $\phi(x,y) = 10^5\ \mathrm{W m^{-2}}$ 
\end{itemize}

\noindent constant in time. The contact resistance between the slabs is due to a variable thickness $d(x,y)$. Given the thermal conductivity $k_a$ of the material between the two sheets (air) 
\begin{itemize}
\item $k_a = 25.87 \times 10^{-3}\ \mathrm{W m^{-1} K^{-1}}$
\end{itemize}

\noindent the unknown heat exchange coefficient is given by 
$$H(x,y) = d(x,y)/k_a .$$\\

For the simulations we assume:

\begin{equation}
d(x,y) = d_0\left(1+ 20 \chi_{E_1}(x,y) + \chi_{E_2}(x,y) + 10 \chi_{E_3}(x,y) + 3 \chi_{E_4}(x,y) \right)
\end{equation}

\noindent
where:

$$E_1 = \left\{x \in \left[\frac{L}{4}-\frac{L}{20},\frac{L}{4}+\frac{L}{20}\right], y \in \left[\frac{L}{4}-\frac{L}{20},\frac{L}{4}+\frac{L}{20}\right] \right\}$$ 

$$E_2 = \left\{x \in \left[\frac{L}{4}-\frac{L}{20},\frac{L}{4}+\frac{L}{20}\right], y \in \left[\frac{3 L}{4}-\frac{L}{40},\frac{3 L}{4}+\frac{L}{40}\right] \right\}$$ 

$$E_3 = \left\{x \in \left[\frac{3 L}{4}-\frac{L}{20},\frac{3 L}{4}+\frac{L}{20}\right], y \in \left[\frac{3 L}{4}-\frac{L}{20},\frac{3 L}{4}+\frac{L}{20}\right] \right\}$$ 

$$E_4 = \left\{x \in \left[\frac{3 L}{4}-\frac{L}{40},\frac{3 L}{4}+\frac{L}{40}\right], y \in \left[\frac{L}{4}-\frac{L}{20},\frac{L}{4}+\frac{L}{20}\right] \right\}$$

\begin{figure}[!ht]
\begin{center}
\includegraphics[width=0.8\textwidth]{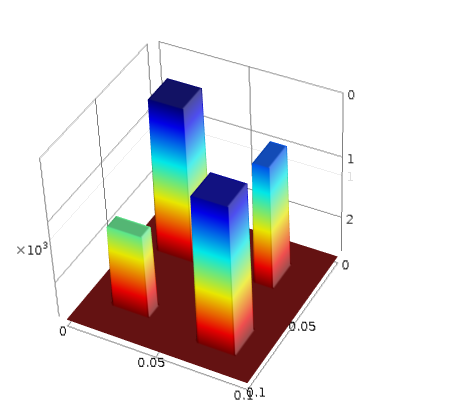}
\caption{Unknown heat exchange coefficient at the interface}
\label{Hinterf}
\end{center}
\end{figure}

\begin{figure}[!ht]
\begin{center}
\includegraphics[width=0.8\textwidth]{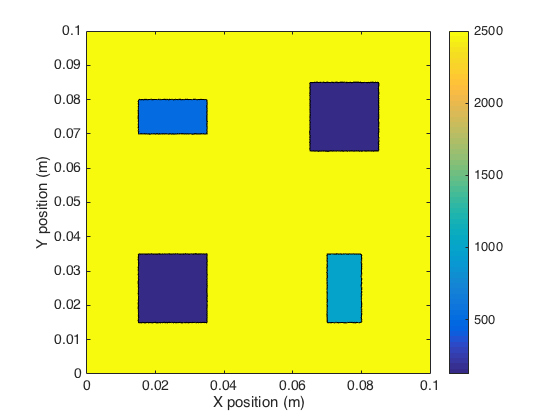}
\caption{Unknown heat exchange coefficient at the interface}
\label{Hinterf1}
\end{center}
\end{figure}

Figure \ref{Hinterf} shows the graph of $H(x,y)$. Figure \ref{Hinterf1} represents  a filled contour plot of $H(x,y)$.

Simulation has been coded in COMSOL Multiphysics. The model is three dimensional,  with the imperfect contact between the two slabs modeled as a  resistive thin layer. The continuous uniform heating is supplied for 100 seconds.

\begin{figure}[!ht]
\begin{center}
\includegraphics[width=0.8\textwidth]{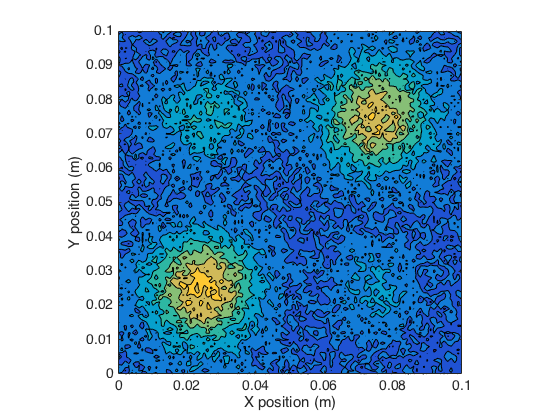}
\caption{Temperature distribution on the accessible surface}
\label{thermo}
\end{center}
\end{figure}

Figure \ref{thermo} shows the temperature distribution ``recorded'' after 50 seconds. A gaussian noise with $\sigma\ \mathrm{=\ 0.1\ ^oC}$ has been added to the simulated temperature map.

\begin{figure}[!ht]
\begin{center}
\includegraphics[width=0.8\textwidth]{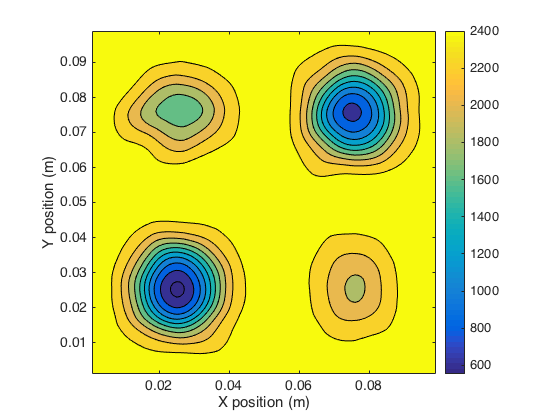}
\caption{Reconstructed $H$ without noise}
\label{Hcalc1}
\end{center}
\end{figure} 


\begin{figure}[!ht]
\begin{center}
\includegraphics[width=0.8\textwidth]{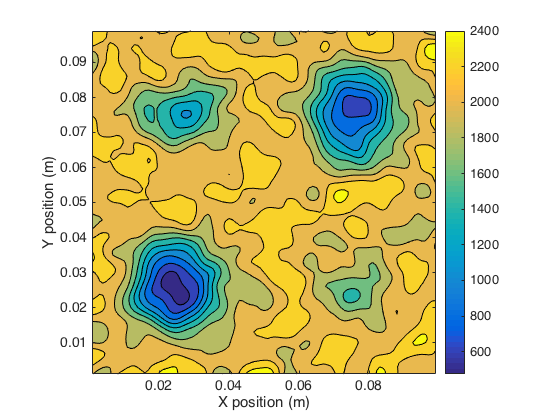}
\caption{Reconstructed $H$ with noise $\sigma$ = 0.1 $^oC$}
\label{Hcalc1_err}
\end{center}
\end{figure} 

\subsection{Recovering $H$ from thermal data collected on the top face of the specimen}

Figures \ref{Hcalc1} and \ref{Hcalc1_err} show the reconstruction of $H(x,y)$ obtained by means of TPA expression (\ref{h_formula}) from exact and noisy data, respectively. In order to compute the spatial derivatives appearing in (\ref{h_formula}), a cubic smoothing spline is applied to data when computing the first-order derivatives. 
Moreover, the same kind of smoothing is applied to temperature maps when noise is present. The weight factors applied to obtain the result shown in figure \ref{Hcalc1_err} are somewhat arbitrary. Since  $H$  is actually not known, the choice of such factors  does not appear to be a straightforward task. 

\subsubsection{Filtering noisy data}

An alternative to using cubic smoothing splines is to filter  noisy data by means of  Fast Fourier Transform (FFT). The following procedure is applied to a temperature  map $u^i$ taken at time $t_i$:
\begin{enumerate}
\item From the spectrum $F_i = F\{u^i\} = FFT(u^i)$, the power spectrum $P_i = |F_i|^2$ is computed. 
\item The spectrum is shifted to the center of the data, and  scaled to have a maximum value 0: $SP_i = log(P_i) - max \left( log(P_i) \right)$.
\item Looking at the graph of the scaled power spectrum (see, for example, figure \ref{spectrum}), a suitable threshold $S_t$ is chosen to cut the image clutter. With such a threshold a mask $M$ is computed (having 0 for $SP_i < S_t$) and multiplied by $F_i$.
\item The filtered temperature map at time $t_i$ is obtained by an inverse Fourier transform: $u_{F}^i = | FFT^{-1}(F_i \cdot \mathrm{M}) |$
\end{enumerate}

\begin{figure}[!ht]
\begin{center}
\includegraphics[width=0.8\textwidth]{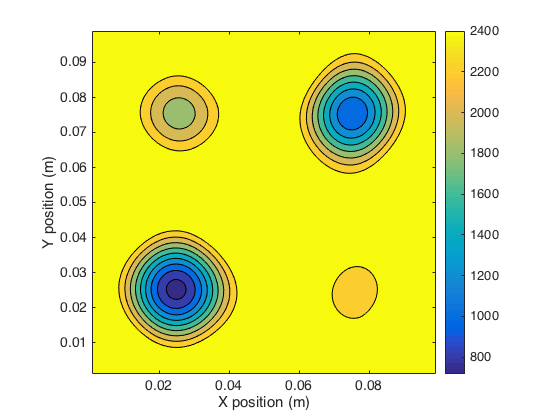}
\caption{Reconstructed $H$ with noise $\sigma$ = 0.1 $^oC$ using FFT filtering}
\label{H_fft}
\end{center}
\end{figure}

The effective heat exchange coefficient computed on the Fourier-filtered map gives the result shown in figure \ref{H_fft}.

\subsubsection{Choice of the filter threshold}
In order to obtain a smooth temperature map from the noisy data, a low-pass filter (LPF) is applied in the frequency domain. The LPF consists in applying a mask to the spectrum based on a suitable threshold $\theta$.
Denoting by $U = F\{u\}$ the Fourier transform of the temperature $u$, and by $S_U = \frac{|U|^2}{max(|U|^2)}$ the normalized power spectrum, the mask $M$ is defined by:

\begin{equation}
M = \left\{
    \begin{array}{rl}
      1 & \text{if } S_U > -\theta,\\
      0 & \text{if } S_U \le -\theta
    \end{array} \right.
\end{equation}

A ``denoised'' map $u_D$ is obtained by inverse-transforming the product $U\cdot M$: $u_D = F^{-1}(U\cdot M)$.The choice of the threshold is not a straightforward task: a small threshold leaves data unmodified, while a too large one smooths data too much.
We adopt the following procedure, which allows an automatic computation of the threshold. If we choose a random value of the threshold, between $min(S_U)$ and $0$, an high-pass filter (HPF) is realized by the mask $M^c$ complementary to $M$. We shall call ``noise'' temperature that obtained by: $u_N = F^{-1}(U \cdot M^c)$

Choosing $\theta = 0$ the HPF gives a perfect replica of the map $u$, while the LPF gives a null power spectrum. Conversely, the LPF gives a perfect (unsmoothed) replica of the temperature data if $\theta = \text{min}(S_U)$ while the HPF gives a null spectrum in such conditions. A small value of $\theta > \text{min}(S_U)$ produces masks $M$ and $M^c$ such to give an almost random temperature.
Therefore, intuitively we should increase the value of $\theta$, starting from the lower value of the power spectrum, until the noise temperature $u_N$ vaguely starts to look like the original map $u$. In those conditions, the map $u_D$ is over-smoothed.

\begin{figure}[!ht]
\begin{center}
\includegraphics[width=0.8\textwidth]{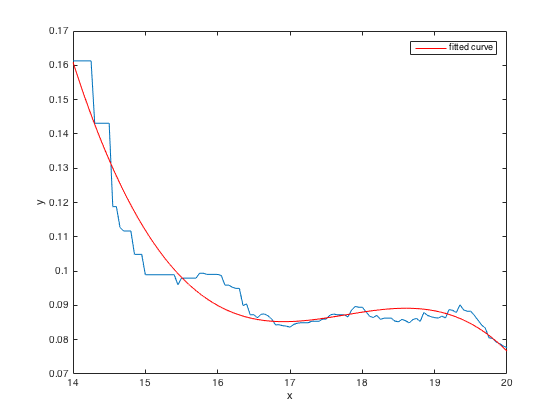}
\caption{Example of correlation $c_{n,s}$}
\label{correlation}
\end{center}
\end{figure}

The previous reasoning can be translated into formulae, by considering what ``to look like'' means in this context. If we compute the correlation among the noise ($n$) temperature map $u_N$ and the original map (signal, $s$) $u$, $c_{n,s}\text{corr}(u_N,u)$, for values of $\theta \in (\text{min}(S_U),0)$, the plot of $c_{n,s}$ versus $\theta$ is like that shown in figure \ref{correlation}, where a third-order polynomial fitting is superimposed to the computed correlation.

The threshold $\theta_0$ corresponding to the first minimum of the fitting curve is such to give the best low-pass filtered data, without ``loosing'' information (i.e. with a noise map $u_N$ actually representing noise).

Looking at figure \ref{correlation} it is clear that such a minimum is not the absolute minimum of the correlation. Clearly, a zero noise temperature is the absolutely less correlated to any non-zero map!

\begin{figure}[!ht]
\begin{center}
\includegraphics[width=0.8\textwidth]{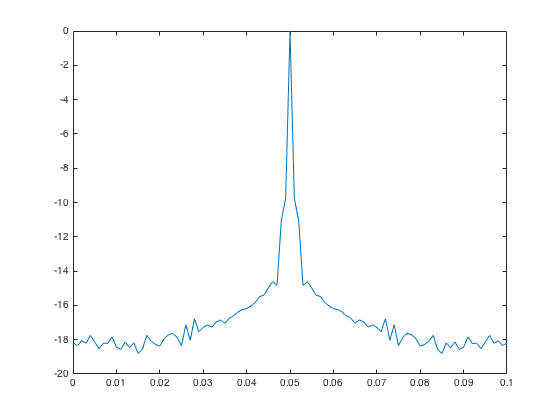}
\caption{Scaled power spectrum for data corresponding to figure \ref{correlation}}
\label{spectrum}
\end{center}
\end{figure}

Figure \ref{spectrum} shows the spectrum corresponding to figure \ref{correlation}, actually a cut of the 3d-spectrum on a symmetry plane.

\subsubsection{Resolution limits}
\begin{figure}[!ht]
\begin{center}
\includegraphics[width=0.8\textwidth]{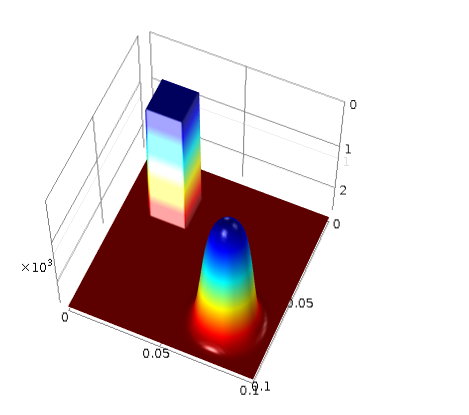}
\caption{Contact resistance profile consisting of a rectangular-shaped and a gaussian $H$}
\label{H3D_conf}
\end{center}
\end{figure}

Gaussian blurring is intrinsic in heat diffusion processes, so we can expect that different contact resistance profiles conduct to similar $H$ shapes. That's indeed the case. Figure \ref{H3D_conf} shows a $H$ profile having two deviations from the base value, respectively with rectangular and gaussian shapes.

\begin{figure}[!ht]
\begin{center}
\includegraphics[width=0.8\textwidth]{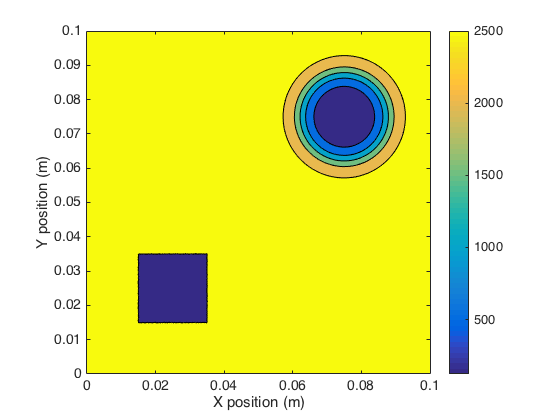}
\caption{Contour plot of the true $H$}
\label{Hvero_conf}
\end{center}
\end{figure}

\begin{figure}[!ht]
\begin{center}
\includegraphics[width=0.8\textwidth]{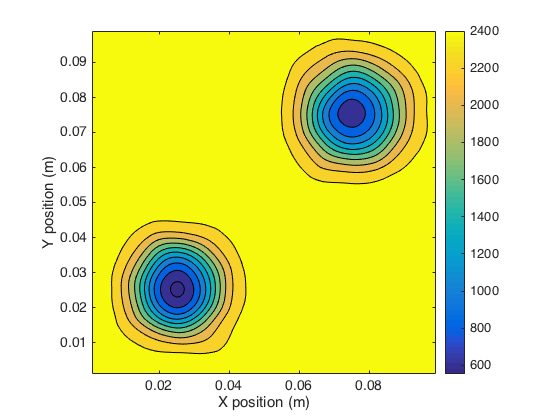}
\caption{Contour plot of the reconstructed $H$ in the noiseless case}
\label{Hcalc_conf}
\end{center}
\end{figure}

\begin{figure}[!ht]
\begin{center}
\includegraphics[width=0.8\textwidth]{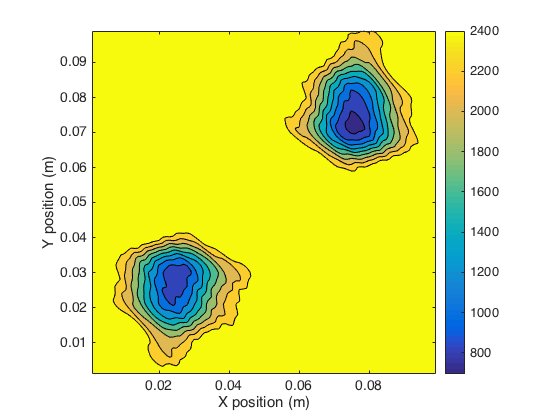}
\caption{Contour plot of the reconstructed $H$ in the noisy case}
\label{Hcut_conf}
\end{center}
\end{figure}

Figures \ref{Hvero_conf} shows the contour plot of the true $H$, to be compared to the reconstructed $H$, without ``measurement'' noise (figure \ref{Hcalc_conf}) and in presence of noise (figure \ref{Hcut_conf}), respectively.

\begin{figure}[!ht]
\begin{center}
\includegraphics[width=0.8\textwidth]{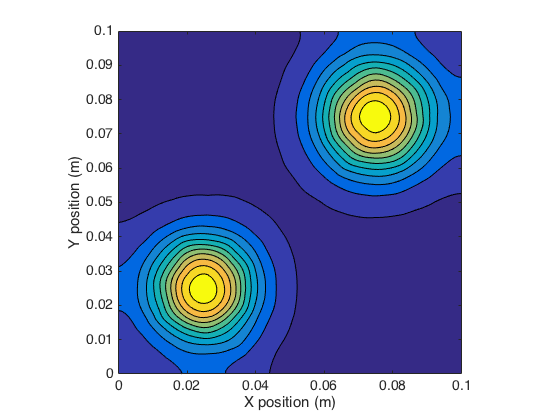}
\caption{True temperature after 10 seconds heating, noiseless case}
\label{Tvera_conf}
\end{center}
\end{figure}

\begin{figure}[!ht]
\begin{center}
\includegraphics[width=0.8\textwidth]{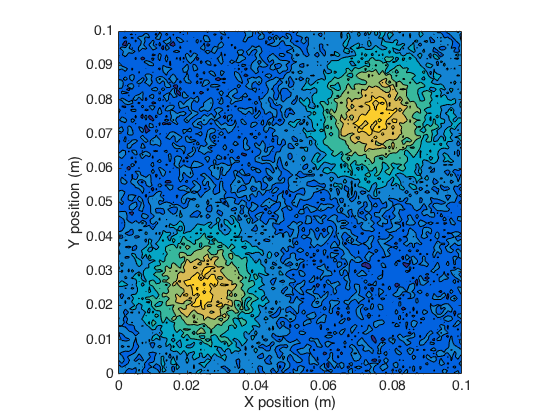}
\caption{True temperature after 10 seconds heating, noisy case}
\label{TveraN_conf}
\end{center}
\end{figure}

As figures \ref{Hcalc_conf} and \ref{Hcut_conf} show, the rectangular and gaussian profiles bring to nearly identical $H$ shapes, as a consequence of temperature blurring, as figures \ref{Tvera_conf} and \ref{TveraN_conf} show, displaying the temperature distributions after 10 seconds heating.

\section{Conclusions} 
Consider a composite body made up of two slabs of different materials in close imperfect thermal contact.
Large deviations of the interface width from a known average value $d_0$ must be detected. The body is heated from above (with a spotlight) while temperature maps of the upper side of the body are collected (with an infrared camera). Since transmission conditions through the interface are transformed into Robin Boundary Conditions, we reduce the evaluation of flaws in the interface to an inverse problem for the heat equation in the upper slab only (section \ref{sec:TrasRob}). The unknown is the heat transfer coefficient $2H(x,y)$ in the  Robin condition where $H$ is the thermal conductance of the interface. A reliable evaluation of $H$ is obtained in section \ref{sec:numer} by means of Thin Plate Approximation and Fast Fourier Transform with simulated data corresponding to a realistic physical situation.

\end{document}